# Shock recovery with decaying compressive pulses: Shock effects in calcite (CaCO$_3$) around the Hugoniot elastic limit

Kosuke Kurosawa[a], Haruka Ono[a], Takafumi Niihara[b,c], Tatsuhiro Sakaiya[d], Tadashi Kondo[d], Naotaka Tomioka[e], Takashi Mikouchi[f], Hidenori Genda[g], Takuya Matsuzaki[h], Masahiro Kayama[i], Mizuho Koike[j], Yuji Sano[h,k], Masafumi Murayama[h], Wataru Satake[l] and Takafumi Matsui[a,l]

[a]Planetary Exploration Research Center, Chiba Institute of Technology, 2-17-1, Tsudanuma, Narashino, Chiba 275-0016, Japan

[b]Department of Systems Innovation, School of Engineering, The University of Tokyo, 7-3-1 Hongo, Bunkyo-ku, Tokyo 113-8656, Japan.

[c]Department of Applied Science, Okayama University of Science, 1-1 Ridaicho, Kita-ku, Okayama, Okayama 113-8656, Japan.

[d]Department of Earth and Space Science, Graduate School of Science, Osaka University, 1-1 Machikaneyama, Toyonaka, Osaka 560-0043, Japan

[e]Kochi Institute for Core Sample Research, Japan Agency for Marine-Earth Science and Technology (JAMSTEC), 200 Monobe Otsu, Nankoku, X-star, Kochi 783-8502, Japan.

[f]The University Museum, The University of Tokyo, 7-3-1 Hongo, Bunkyo-ku, Tokyo 113-0033, Japan.

[g]Earth–Life Science Institute, Tokyo Institute of Technology, 2-12-1 Ookayama, Meguro-ku, Tokyo 152-8550, Japan.

[h]Center for Advanced Marine Core Research, Kochi University, 200 Monobe Otsu, Nankoku, Kochi 783-8502, Japan.

[i]Department of General Systems Studies, Graduate School of Arts and Sciences, The University of Tokyo, 3-8-1, Komaba, Meguro-ku, Tokyo 153-8902, Japan

[j]Department of Earth and Planetary Systems Science, Graduate School of Science, Hiroshima University, 1-3-1, Kagamiyama, Higashi–Hiroshima, Hiroshima 739-8526, Japan

[k]Atmosphere and Ocean Research Institute, The University of Tokyo, 5-1-5, Kashiwanoha, Kashiwa, Chiba 277-8564, Japan

[l]Institute of Geo-Cosmology, Chiba Institute of Technology, 2-17-1, Tsudanuma, Narashino, Chiba 275-0016, Japan

*Corresponding author
Kosuke Kurosawa PhD
Planetary Exploration Research Center, Chiba Institute of Technology
E-mail: kosuke.kurosawa@perc.it-chiba.ac.jp

**Abstract**

Shock metamorphism of minerals in meteorites provides insights into the ancient Solar System. Calcite is an abundant aqueous alteration mineral in carbonaceous chondrites. Return samples from the asteroids Ryugu and Bennu are expected to contain calcite-group minerals. Although shock metamorphism in silicates has been well studied, such data for aqueous alteration minerals are limited. Here, we investigated the shock effects in calcite with marble using impact experiments at the Planetary Exploration Research Center of Chiba Institute of Technology. We produced decaying compressive pulses with a smaller projectile than the target. A metal container facilitates recovery of a sample that retains its pre-impact stratigraphy. We estimated the peak pressure distributions in the samples with the iSALE shock physics code. The capability of this method to produce shocked grains that have experienced different degrees of metamorphism from a single experiment is an advantage over conventional uniaxial shock recovery experiments. The shocked samples were investigated by polarizing microscopy and X-ray diffraction analysis. We found that more than half of calcite grains exhibit undulatory extinction when peak pressure exceeds 3 GPa. This shock pressure is one order of magnitude higher than the Hugoniot elastic limit (HEL) of marble, but it is close to the HEL of a calcite crystal, suggesting that the undulatory extinction records dislocation-induced plastic deformation in the crystal. Finally, we propose a strategy to re-construct the maximum depth of calcite grains in a meteorite parent body, if shocked calcite grains are identified in chondrites and/or return samples from Ryugu and Bennu.

**Introduction**

Shock metamorphism recorded in chondrites has been used to reconstruct the impact histories of their parent bodies [e.g., Stöffler et al., 1991, 2018]. Shock effects for silicates in meteorites have been categorized into progressive stages based on uniaxial shock recovery experiments [e.g., Stöffler et al., 1991, 2018]. However, primary silicates were sometimes metamorphosed after aqueous alteration in parent bodies of chondrites, which resulted in formation of hydrous minerals and aqueous alteration [e.g., Krot et al., 2015; Rubin et al., 2007]. Given that the shock effects in hydrous and aqueous alteration minerals are well understood, it may be possible to constrain the depth below the surface of water–rock–organic reactions within meteorite parent bodies, which also produce a variety of complex organic species [e.g., Lindgren et al., 2013]. In this study, we focused on the shock metamorphism of calcite ($CaCO_3$), which is an aqueous alteration mineral, because calcite and other carbonate group minerals (i.e., siderite [$FeCO_3$] and dolomite [$CaMg(CO_3)_2$]) have been identified in hydrated carbonaceous chondrites [e.g., King et al., 2017]. In addition, veins containing carbonates were found on the surface of the carbonaceous asteroid Bennu [Kaplan et al., 2020]. As such, calcite-group minerals are expected to be present in the return samples from Ryugu and Bennu.

The production of mechanical twins [e.g., Lindgren et al., 2013] and impact devolatilization [e.g.,

Boslough et al., 1982; Lange and Ahrens, 1986; Ohno et al., 2008; Kurosawa et al., 2012; Bell, 2016] are known shock effects of calcite [e.g., Langenhorst, 2002]. In addition, crystal lattice dislocation in shocked calcite samples has been identified by transmission electron microscopy [e.g., Barbar and Wenk, 1979; Bell, 2016]. The required peak pressures for twin production and incipient devolatilization have been estimated to be 0.5 GPa [Lindgren et al., 2013] and 20 GPa [e.g., Kurosawa et al., 2012, 2021a], respectively. Such shock features in calcite are expected to be a powerful tool in constraining the depth of aqueous alteration in the parent bodies of carbonaceous chondrites [e.g., Lindgren et al., 2013]. For example, given that the experienced shock pressure is known from the petrographic features, we can estimate the distance from the impact point (i.e., the surface), because impact-induced compressive waves rapidly attenuate during their propagation, following a power law with distance from the impact point with an exponent of 2–3 [e.g., Pierazzo et al., 1997]. However, accurate depth estimation is difficult because there is a large gap between the threshold pressures for twin production (<0.5 GPa) and devolatilization (>20 GPa). In this study, we investigated the shock effects in calcite at an intermediate range of shock pressures between 0.5 and 20 GPa.

The remainder of the manuscript is divided into four sections. We describe our methodology in detail in Section 2 The results are presented in Section 3. We found that shocked calcite grains exhibit undulatory extinction as a response to shock pressures, much like several other silicate minerals. In Section 4, we introduce the characteristic peak pressure for producing undulatory extinction in more than half of calcite grains $P_{UE}$ and quantify the $P_{UE}$ pertaining to calcite. Then, we discuss about the strategy to conduct the depth estimation of calcite based on the presence of the shocked calcite grains showing undulatory extinction. We summarize the study in Section 5.

1. **Methodology: impact experiments, analyses of shocked samples, and shock physics modeling**

There are three types of techniques for shock recovery experiments. At the beginning of this section, we discuss the advantages and disadvantages of each technique. The first technique has been referred to as uniaxial shock recovery experiments [e.g., Langenhorst and Hornemann, 2005; Bell, 2016]. This technique has also been termed shock reverberation or ring-up [e.g., Meyers, 1994]. In such experiments, a specimen is shaped into a thin disk and is placed inside a metal container. A flyer plate collides onto a flat surface of the metal container. A planar compressive pulse then propagates into the front plate of the container and the specimen. When the compressive pulse reaches the rear surface of the specimen, the pulse is reflected from the interface between the specimen and container wall, because of the larger shock impedance for metal than that of the rocky specimen. The pressure experienced by the specimen gradually becomes larger for every wave arrival. This method allows the peak pressure to be accurately calculated from the measured flyer velocity and one-dimensional impedance matching method [e.g., Melosh, 1989, pp. 54–57] if the shock Hugoniot parameters for the materials are known. In addition, this method can produce relatively high peak pressures due to multiple compressions. Given that the wave reflection in the closed system of a metal container does not occur in natural impact phenomena, the difference in geometry should be carefully considered when such experimental results are applied to natural samples [e.g., Ivanov and Deutsch, 2002; Sharp and DeCarli, 2006].

The second technique is a standard cratering experiment with a relatively large block. Although the shocked structure deep inside the target can be examined in a fully open system, which is the same geometry as in natural impact events [e.g., Lindgren et al., 2013; Winkler et al., 2018; Ohtani et al., 2022], the highly shocked materials around the impact point are lost due to impact spallation and excavation [e.g., Polanskey and Ahrens, 1990; Suzuki et al., 2021]. The first and second methods are time consuming and expensive to carry out. For the first method, we can only obtain data at one peak pressure per shot, because the entire specimen experiences the same peak pressure. In addition, the metal container is significantly distorted by each shot. The second method requires blocks that are larger than ~10 cm in size to prevent catastrophic disruption of the specimen. For example, Winkler et al. (2018) used a marble cube with 25 cm long sides (i.e., >40 kg in weight). Note that Lindgren et al. (2013) used the second method with marble blocks with the size of a few cm, and quantified the condition for producing mechanical twins to be 0.11–0.48 GPa by combining shock physics modelling. The projectile size and impact velocity were limited to ~1 mm and ~2 km s$^{-1}$, respectively, possibly due to catastrophic disruption of the specimen. Thus, further extension to higher pressures is difficult when we use samples with the size of a few cm in the second method.

The third technique uses decaying compressive pulses and metal containers [e.g., Langenhorst and Hornemann, 2005; Kowitz et al., 2013a, b; Nagaki et al., 2016]. For examples, the decaying compressive pulses have been produced by high explosives [e.g., Kowitz et al., 2013a, b] and high-power lasers [Nagaki et al., 2016]. When the spatial scale of the epicenter or "isobaric core" is much smaller than the specimen size, decaying compressive pulses propagate into the specimen. Although the reflected waves are produced in the same way as the first method, the strengths of the reflected waves are much smaller than for the first method, because of the rapid attenuation of the compressive pulse with increasing distance from the epicenter. As such, it is possible to ignore the effects of the reflected compressive pulses when the size ratio of the epicenter to the specimen is small enough such that the reflected pulses are much weaker than the initial pulse. In addition, it is possible to collect shocked targets that have experienced a range of peak pressures and largely retain their pre-impact stratigraphy, because the peak pressures depend on the initial distance from the epicenter. The disadvantage of the third method is that the peak pressures cannot be calculated analytically. However, we can solve this problem by using numerical calculations as performed by the previous studies [e.g., Lindgren et al., 2013; Nagaki et al., 2016]. Recent shock physics codes can quantify the hydrodynamic and elasto-plastic responses of shocked materials in a self-consistent fashion [e.g., Ivanov et al. 1997; Mitani, 2003], which allows the peak pressure distributions to be accurately estimated as a function of distance from the epicenter [e.g., Mitani, 2003; Nagaki et al., 2016]. Consequently, we used the third method and a two-stage light gas gun. The method has already been applied to laser shock experiments on single olivine crystals [Nagaki et al., 2016]. We applied this method to macro blocks of calcite (i.e., marble). It should be noted that all three methods require some type of equations of state (EOS) model to calculate the peak pressures in the recovered samples. The remainder of this section is divided into three sub-sections. We describe the details of the impact experiments, sample analyses, and shock physics modeling in sections 2.1, 2.2, and 2.3, respectively.

**1.1. Impact experiments**

We conducted a series of impact experiments with a two-stage light gas gun in the Hypervelocity Impact Laboratory of the Planetary Exploration Research Center (PERC) of the Chiba Institute of Technology (Chitech), Japan [Kurosawa et al., 2015]. We produced an expanding (i.e., decaying) shock wave by using projectiles that are much smaller than the metal containers. A polycarbonate sphere with a diameter of 4.8 mm was used as the projectile. We used a titanium container with a diameter of 50 mm and a detachable front plate for shock recovery. A schematic diagram of the experimental set-up is shown in Figure 1. The dimensions of the sample and titanium container are shown in the figure. We only conducted vertical impacts onto a cylinder of Carrara marble. The cylinder had a diameter of 30 mm and length of 24 mm. Carrara marble consists of crystallographically randomly oriented calcite crystals that are 100–300 μm in diameter. We used titanium and aluminum front plates with thicknesses of 3 or 4 mm. The impact velocity was varied from 5.1 to 7.3 km s$^{-1}$. The material and thickness of the front plate and impact velocity are the main factors that determine the peak pressure at the epicenter in the sample. The experimental conditions are summarized in Table 1. We also conducted two shock recovery experiments with stress gauges to directly measure the maximum principal stress at the epicenter and far end of the sample. These results were used to confirm the validity of the shock physics modeling.

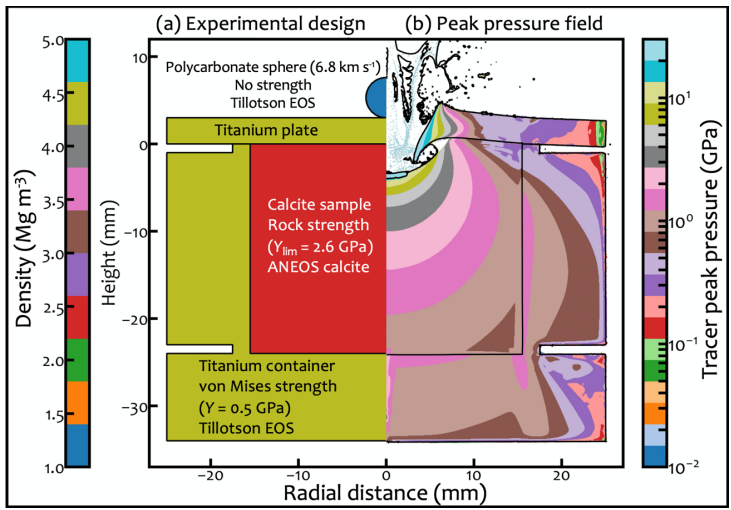

**Figure 1.** (a) The dimensions of the shock recovery experiments shown in a provenance plot. The color scale corresponds to the initial density. The material type, strength model, and Equation of State model used in the simulation are also shown in the figure. (b) The peak pressure field after a compressive wave moves through the entire target. The color scale corresponds to tracer peak pressure and is shown on a log scale.

**Table 1.** Experimental conditions. The mass of the projectile was 67.9 mg for all shots.

| Shot number | Impact velocity (km s$^{-1}$) | Front plate material | Plate thickness (mm) | Stress gauge |
|---|---|---|---|---|
| 470 | 5.48 | Titanium | 4.0 | without |
| 473 | 7.32 | Titanium | 3.0 | without |
| 475 | 6.82 | Titanium | 3.0 | without |
| 486 | 5.96 | Aluminum | 4.0 | with |
| 487 | 5.10 | Aluminum | 4.0 | with |

**2.2. Sample analyses**

We made the collected shocked samples into polished thin-sections. Given that we cut the samples across the centers of the shock-generated depressions on the top surfaces of the shocked samples, the polished thin-sections exhibit cross-sections parallel to the projectile trajectory. We observed the samples with a polarizing microscope, and a micro-X-ray diffraction analyzer (µXRD; Rigaku, RINT RAPID II) was used to analyze one of the shocked samples (shot #475). We also conducted three-dimensional observations of the shocked sample from shot #475 with a micro-focus X-ray computational tomography (XCT) device (Zeiss Xradia 410 Versa) at Kochi University, Japan, prior to making the polished thin-section. We focused on the petrographic shock feature termed undulatory extinction. It is widely known that some shocked silicate crystals exhibit a characteristic extinction pattern under polarized microscopy [e.g., Stöffler et al., 2018]. At a given stage angle under cross-polarized light, different parts of a single grain exhibit extinction. The extinction region then sweeps through the entire grain as the stage of the microscope is rotated. We examined a rectangular region within 1.3 mm of the epicenter in the radial direction (i.e., 2.6 mm wide). The rectangle was divided into ~20 sub-domains with lengths of 1.7 mm, which included ~100 grains. The sub-domains slightly overlap adjacent sub-domains. We chose ~10 grains from each domain and observed these with a 20× objective under the microscope, following the procedure of Stöffler et al. (2018) to examine whether a given grain exhibited undulatory extinction or not. Given that the grain counting method is highly subjective, three observers of the present study independently conducted the measurements. Finally, we calculated the fraction of grains showing undulatory extinction in each sub-domain. It should be noted that the number density of mechanical twins under strong compression would be too high to identify individual mechanical twins by optical microscopy. The extinction pattern of such densely twinned grains is largely indistinguishable from undulatory extinction under a polarizing microscope. We counted such grains as having undulatory extinction in our measurements. This is because both mechanical twins and undulatory extinction form by plastic deformation of the crystal lattice [Blenkinsop, 2007, pp. 41–43].

We also observed the polished thin-sections by µXRD at Osaka University, Japan. The change in the crystallographic state of calcite is recorded by the XRD patterns. We used copper as the X-ray source and optics (Rigaku; VariMax dual-wavelength) to monochromatize and focus the X-ray beam to a spot. The sample table was tilted 45° from the horizontal, and the incidence angle of the X-rays to the sample was 30°. The X-ray spot was an ellipse with a minor axis of 1.1 mm (i.e., the depth direction) and major axis of 1.6 mm (i.e., the radial direction) on the surface of the polished thin-sections. The field of view of the XRD analyses was slightly smaller than the size of the sub-domains in the microscopic measurements. As such, we simultaneously observed >30 calcite grains. The obtained XRD data contain information from a number of calcite grains with different crystal orientations. The exposure time of X-ray was 5 min for each measurement. The X-ray diffraction patterns from the sample were observed in two dimensions on a curved imaging plate. The measurements were conducted at intervals of 1 mm on polished thin-sections from beneath the epicenter to the rear surface of the target. We obtained two-dimensional X-ray diffraction (2D-XRD) patterns, which are the diffraction intensity as a function of the sum of incidence and exit angles ($2\theta$) and another scattering angle $\chi$ [e.g., Izawa et al., 2011; Rupert et al., 2020]. The former provides information about the crystal structure based on Bragg's law. The intensity ratios and the full-widths-of-half-maximum of the diffraction peaks are useful indicators of shock metamorphism [e.g., Bell, 2016; Imae and Kimura, 2021]. The latter provides information about the degree of randomness pertaining to the crystal orientations of calcite grains within the field of view. For example, the 2D-XRD method has been applied to the analysis of powder samples. A ring pattern in the $\chi$ direction can be observed when there are a sufficient number of grains under the X-ray spot, because grains in a powder sample have a range of crystal orientations. The ring structure is referred to as a Debye–Scherrer ring [e.g., Rupert et al., 2020].

**2.3. Shock physics modeling**

We used the most recent stable release of the two-dimensional version of the iSALE shock physics code [Amsden et al., 1980; Ivanov et al., 1997; Wünnemann et al., 2006], iSALE-Dellen [Collins et al., 2016], to estimate the peak pressures in the shocked samples. We used two-dimensional cylindrical coordinates to model the vertical impacts performed in the impact experiments. The projectile in the simulation was divided into 46 cells per projectile radius. This value is high enough to accurately solve for the peak pressure distribution [e.g., Pierazzo et al., 2008]. The dimensions for the projectile, marble sample, and metal container in the simulation were the same as in the impact experiment. We used Tillotson equations of state (EOS) [Tillotson, 1962] for polycarbonate [Sugita and Schultz, 2003] and titanium [Tillotson, 1962], and the analytical EOS (ANEOS) [Thompson and Lauson, 1972] for calcite [Pierazzo et al., 1998]. The "ROCK" [Collins et al., 2004] and von-Mises strength models in the iSALE package [Collins et al., 2016] were used for the marble sample and titanium container, respectively. We also used the von-Mises and Johnson-Cook models [Johnson and Cook, 1983] for the detachable front/rear plates made of titanium and aluminum, respectively. Lagrangian tracer particles were inserted into computational cells to store the temporal variations of pressure. The mass-weighted averages of the peak pressures of the tracers located within the sub-domains defined in the microscopic measurement were calculated. Further details of the shock physics modeling are provided in Supplementary Information Text S1. We also conducted sensitivity tests by changing the input parameters, such as spatial resolution and the strength model parameters, as described in Supplementary Information Texts S2 and S3.

## 3. Results

We first characterize the decaying compressive pulse produced in our laboratory, based on the iSALE modeling in section 3.1. Secondly, we present microscopic observations of the shocked samples in section 3.2. Thirdly, we present the threshold peak pressure for producing undulatory extinction in calcite in section 3.3. Finally, we present the XRD spectra as a function of distance from the epicenter in section 3.4.

### 3.1. Characterization of the decaying compressive pulses

We calculated the propagation behavior of the generated compressive pulse due to a projectile collision onto a front plate with iSALE. Figure 2 shows snapshots of temporal values, including pressure, temperature, volumetric strain, and strain rate. In this figure, we show the calculation corresponding to shot #475. The pressure fields shown in Fig. 2 indicate that the strength of the reflected compressive pulse from the sidewall of the titanium container is much weaker than that of the first pulse. The residual temperature ($T < 800$ K) is not high enough to cause devolatilization of calcite (~1,200 K at $10^5$ Pa [Ivanov and Deutsch, 2002]), even at the epicenter. The volumetric strain of the shocked marble at the epicenter reaches –0.3. The strain rates during compression at the epicenter and far end of the sample are estimated to be ~$10^6$ and ~$10^5$ s$^{-1}$, respectively. Note that the absolute value of the strain rate in the simulations strongly controlled by the artificial viscosity parameters and the spatial resolution (See Supplementary Text S2). Here, we demonstrated that the degree of the spatial variation of the strain rate is about one order of magnitude. Figure 3 shows the cumulative masses experiencing pressures and temperatures higher than the given peak pressure and peak temperature, respectively. The maximum values of the peak pressure and peak temperature are estimated to be 13 GPa and 750 K, respectively.

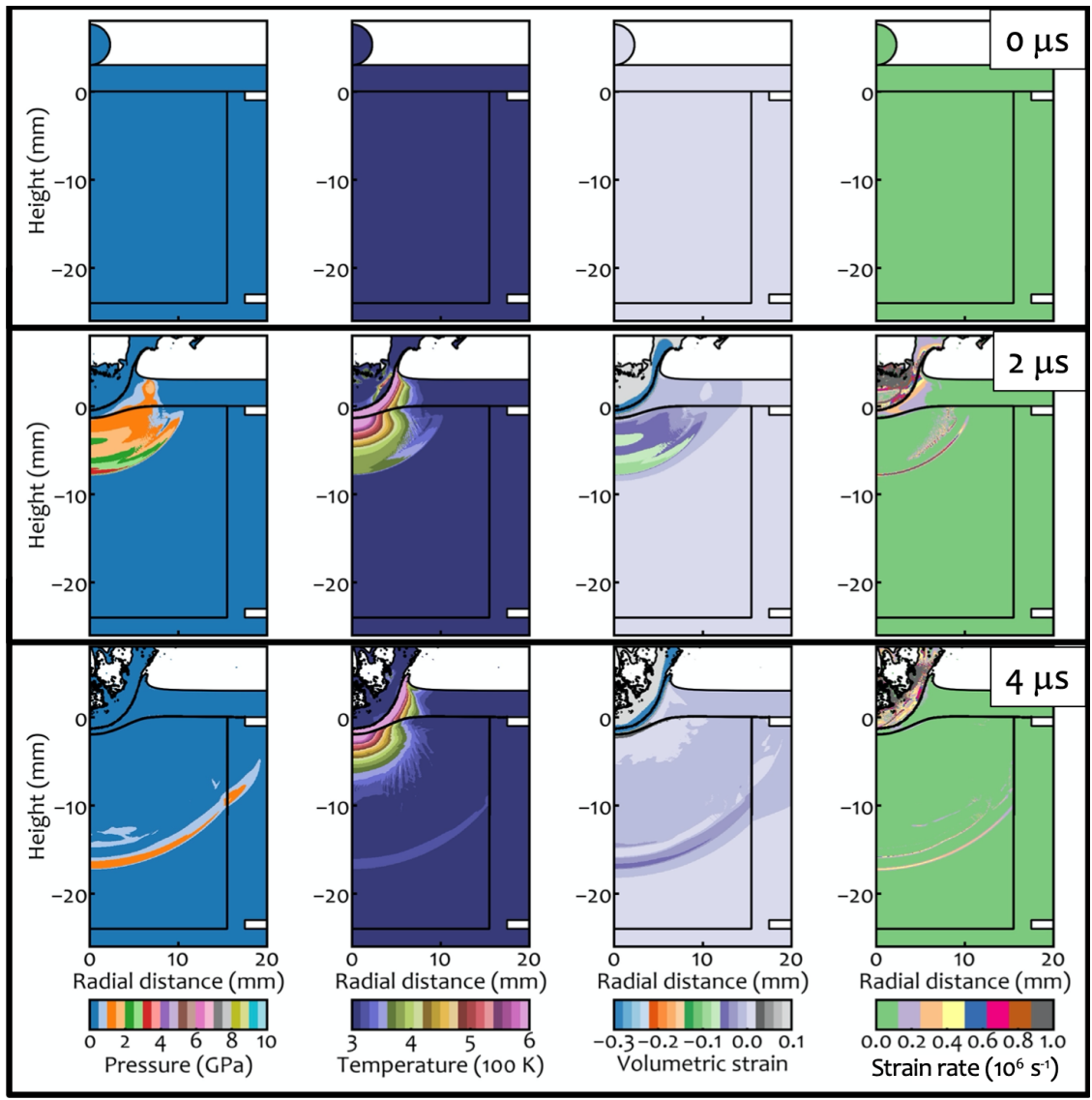

**Figure 2.** Snapshots of the iSALE simulation. The impact velocity is the same as for shot #475 (6.82 km s$^{-1}$). The four different fields (temporal changes in pressure, temperature, volumetric strain, and strain rate) are shown with different color scales. The times in the simulation are also shown.

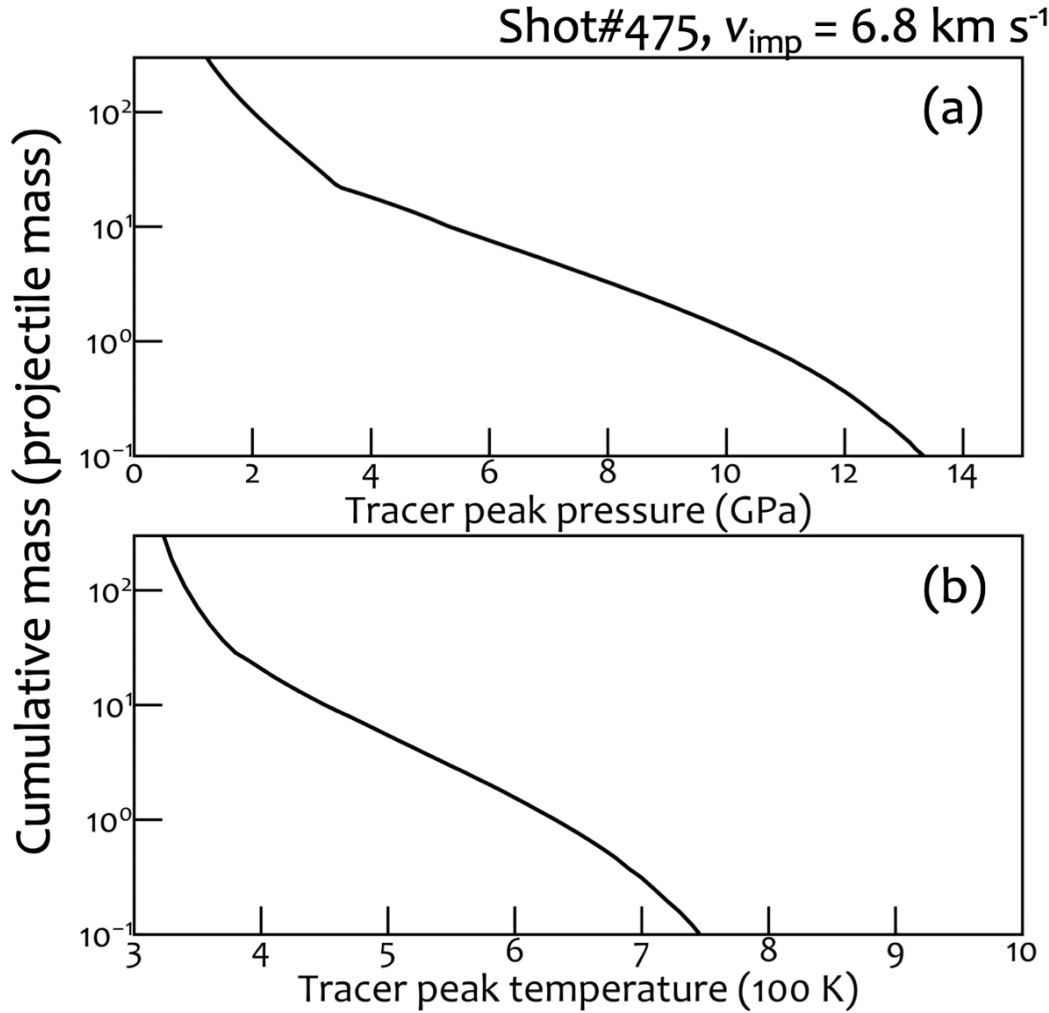

**Figure 3.** (a) Cumulative mass experiencing a peak pressure higher than the given peak pressure as a function of peak pressure for shot #475. (b) Same as (a), except that the cumulative mass for peak temperature is shown. The y-axis is a log scale.

The validity of the iSALE modeling was investigated with stress gauges. We inserted two stress gauges into the boundaries between the front/rear surfaces of the sample and detachable front/rear aluminum plates. We confirmed that the maximum principal stress $\sigma_L$ at the epicenter (i.e., recorded by the front gauge) calculated by iSALE is consistent with the in situ stress measurements. At the rear gauge, we clearly detected wave splitting, which is often called an elastic precursor [e.g., Melosh, 1989, pp. 36–37]. Given that wave splitting is not accounted for with the ANEOS input parameters of the calcite EOS [Pierazzo et al., 1998] used in this study, iSALE tends to overestimate the stress near the rear surface. The degree of overestimation of the peak pressure is likely to increase with distance from the epicenter. However, the difference in $\sigma_L$ in the target between the experiment and simulation is within a factor of two, even at the rear surface of the target, where the difference is expected to be the largest. The details of the stress measurements are presented in Supplementary Information Text S4.

### 3.2. Optical microscopy

We confirmed that the damage structure in the shocked sample has no clear dependence on the azimuth angle with respect to the impact direction by using the 3D-XCT results (Supplementary Information Text S5). The polished thin-sections were observed under an optical microscope with crossed polars. Figure 4a shows a transmitted light image of the whole thin-section for shot #475 under cross-polarized light. Figure 4b shows the peak pressure distribution within the white rectangle as estimated by iSALE. We collected the shocked sample from a region that experienced pressures of 1–10 GPa, which almost retains its initial stratigraphy, although the marble initially located around the front plate (i.e., above the white dotted line in the figure) was comminuted and mixed with other grains from unknown initial locations. The peak pressure for each grain depends on the distance from the epicenter. The red dotted rectangle on the white rectangle in Fig. 4a is an example of a sub-domain. Note that a reduction in transmittance near the epicenter was observed as shown in Fig. 4.

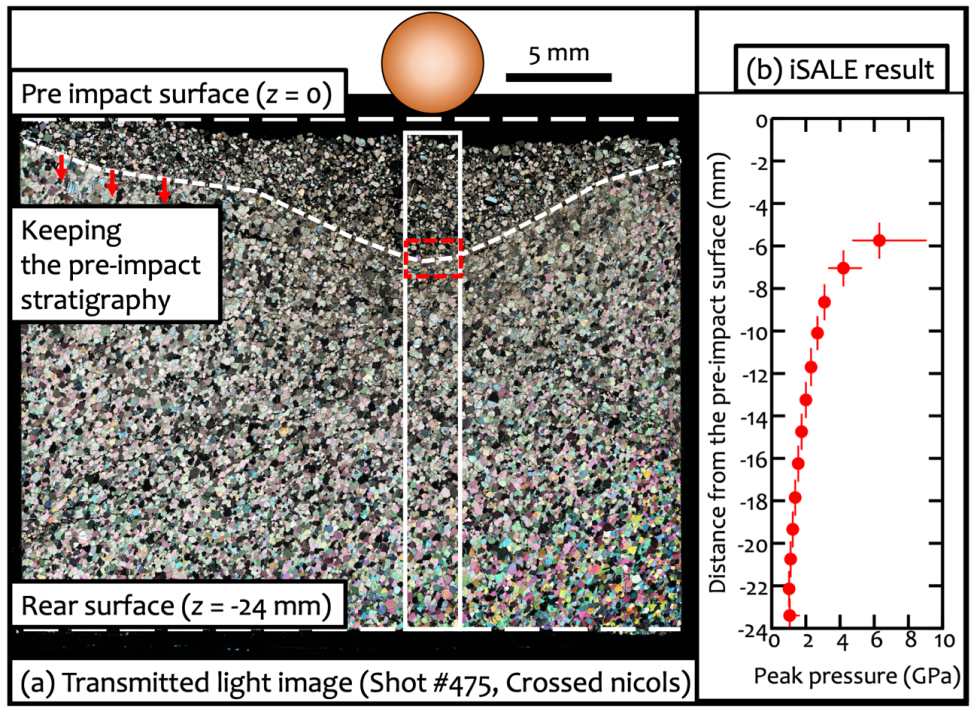

**Figure 4.** (a) A transmitted light image of the thin-section of the shocked sample from shot #475 under cross-polarized light. The white rectangle indicates the region investigated with a polarizing microscope. The red rectangle on the white rectangle corresponds to the size of a sub-domain. (b) The mass-weighted averages of the peak pressure in each sub-domain as a function of the distance from the pre-impact surface. The error bars show the highest and lowest values of the peak pressures stored by the tracer particles located in each sub-domain.

Figures 5a and 5b show transmitted light images of the polished thin-section for shot #486 under cross-polarized light. The size of the observed region was the same as the red rectangle shown in Fig. 4a. The images in Fig. 5a–b were taken at two different locations ($z = -20$ and $-5$ mm, respectively). Figure 5c shows an enlarged image of the rectangular region in Fig. 5a. We indicate the locations of the cracks with different sizes as "micro" and "macro" cracks. For comparison, we also present a transmitted light image of intact Carrara marble in Fig. 5d. Although a few tens of percent of the calcite grains in our intact sample have mechanical twins (Figure 5d), almost all the shocked calcite grains (Figure 5b) have mechanical twins. Thus, the production of mechanical twins even occurred at the far end of the sample ($z = -20$ mm). In addition, some fractures were also observed in some grains. The transmitted light image near the epicenter ($z = -5$ mm; Fig. 5a) is quite different from those of the grains located far from the epicenter. The calcite grains are severely damaged. The reduction in transmittance near the epicenter shown in Fig. 4 is possibly due to a number of micro-fractures in calcite grains. We also observed open cracks between grains that occur as linear voids. The open micro-cracks were also developed between individual grains. The open micro-cracks might contribute to porosity enhancement during impact processes. This result is consistent with a previous shock recovery experiment performed in an open system with a large marble block [Winkler et al., 2018].

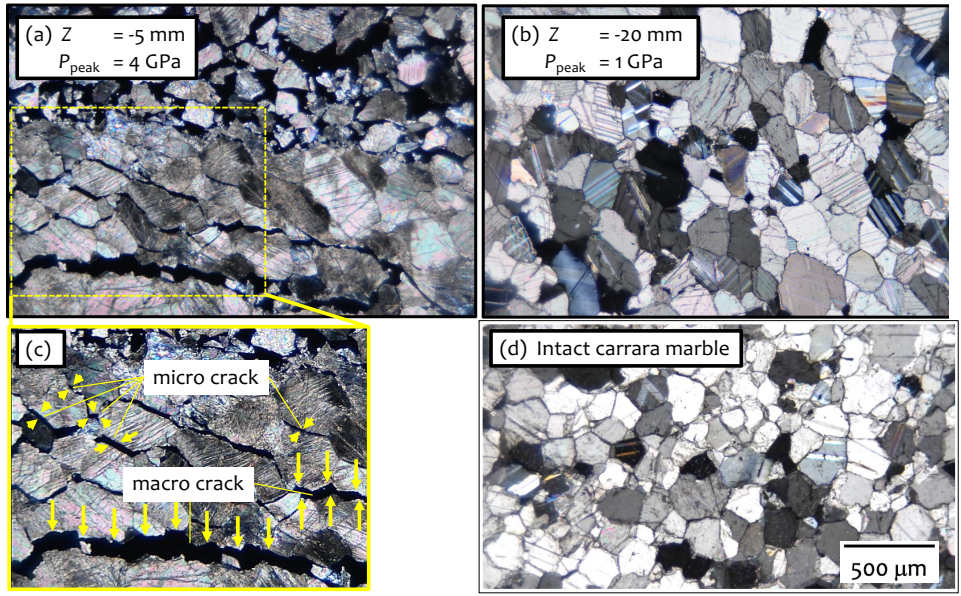

**Figure 5.** Transmitted light images of the (a–c) shocked and (d) intact samples under cross-polarized light. The images were taken of the polished thin-section from shot #486. The locations and estimated peak pressures $P_{peak}$ of (a) and (b) correspond to $z = -5$ mm (4 GPa) and $z = -20$ mm (1 GPa), respectively, where $z$ is the distance from the epicenter in a direction parallel to the projectile trajectory. An enlarged image of the rectangular region in the panel (a) is shown in the panel (c). We indicate the locations of micro and macro cracks with yellow arrows. The spatial scale is shown in the panel (d).

### 3.3. Undulatory extinction in calcite grains

The most prominent metamorphic feature in the shocked samples, especially near the epicenter, is undulatory extinction in the calcite grains, which is a known shock feature in silicate minerals [e.g., Stöffler et al., 2018]. We found that calcite grains showing undulatory extinction were produced in all the shocked samples, and that the number of such grains decreases with increasing distance from the epicenter. In addition, we confirmed that the calcite grains in the polished thin-section of the intact sample do not exhibit undulatory extinction, clearly showing that this can be used as a shock barometer for calcite. As such, we quantified the number fraction of the grains showing undulatory extinction (UEF) in each sub-domain as described in section 2.2. Figure 6a shows UEF as a function of the mass-weighted averages of the peak pressures $P_{peak}$ of the sub-domains estimated by iSALE. We obtained more than 200 UEF data points (5 shots × 15–20 sub-domains × 3 observers) from $P_{peak}$ = 0.7–15 GPa from the five impact experiments. This efficient means of data accumulation is one of the most important advantages of our experimental method. Although the data have a large scatter due to the subjective nature of the UEF measurements, we derived a probability density function (PDF) for a range of $P_{peak}$ values by statistical analysis. The details of the statistical analysis are as follows. Firstly, we defined seven peak pressure bins from 1–4 GPa with steps of 0.5 GPa, except that the bin having the lowest peak pressure also includes all peak pressures of <1 GPa. Secondly, we produced a subset of the UEF data by randomly taking half of the data points from the original dataset by taking the uncertainties in $P_{peak}$ into account. Each data point has a UEF value and corresponding peak pressure estimated from iSALE. Thirdly, we assigned the data points in the sub-dataset to the corresponding bins of the peak pressures. Fourthly, we calculated the average UEF values in the pressure bins. Finally, we repeated the above four steps 3,000 times, and calculated the histogram of the UEF values for each pressure bin. Figure 6b shows the PDFs for $P_{peak}$ bins. We confirmed that UEF > 0.5 at $P_{peak}$ > 2.5–3.0 GPa at the 99.9% confidence level.

### 3.4. X-ray diffraction analysis

We obtained XRD spectra for the selected region in the polished thin-section from shot #475. As mentioned in section 2.2, we obtained 2D XRD spectra. Figure 7 shows the raw data of the 2D XRD image on a curved imaging plate, which were obtained from the polished thin-section for shot #475 at $z$ = –6.7 mm. This area corresponds to the location with UEF >0.5. We confirmed that most of the identified diffraction peaks correspond to calcite. We clearly observed a ring-like pattern in the raw data. We remapped these raw data on a $\chi$ –$2\theta$ plane. Figure 8a–f shows six re-mapped 2D XRD spectra at different locations. Figure 8g is the same as Fig. 4b, except that the locations of the XRD data shown in Fig. 8a–f are shown as blue shaded regions. The two XRD spectra (Fig. 8a–b) are for calcite grains that experienced a peak pressure higher than 3 GPa, which is the characteristic pressure for producing undulatory extinction in more than half of shocked calcite grains. We clearly observed a line in the $\chi$ direction, and no obvious peak broadening in the $2\theta$ direction. The other four XRD spectra (Fig. 8c–f) were obtained for calcite grains with

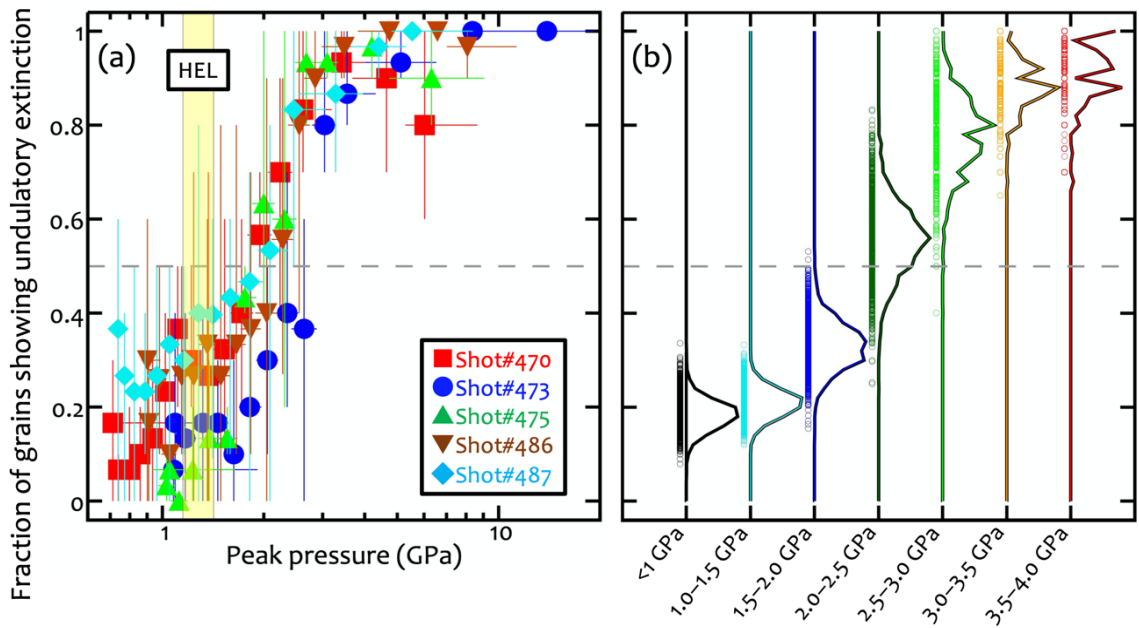

**Figure 6.** (a) The fraction of grains showing undulatory extinction (UEF) as a function of peak pressure. The UEF data were obtained by three different researchers. The average, maximum, and minimum values are plotted as symbols and the upper and lower limits of error bars, respectively. The corresponding peak pressures for the data were calculated with iSALE. The UEF data obtained from five shots are plotted with different colors and symbols as shown in the legend. The horizontal dashed line corresponds to UEF = 0.5. The yellow shaded region is the mean stress at the Hugoniot Elastic Limit (HEL) for calcite crystal [Ahrens and Gregson, 1964]. (b) The histograms of the UEF values at a given bin of peak pressure. The data scatter in a bin is also shown beside the line. The bin size of the UEF values was set to 0.02.

UEF < 0.5. The XRD patterns in Fig. 8c–f exhibit more spotty structures than those in Fig. 8a–b. The broad structures in the $\chi$ direction in Fig. 8a–b suggest that the randomness of the crystal orientations of the calcite grains within the X-ray spot increased due to the compressive pulses. Although further quantitative analyses are needed to confirm this conclusion, we did not observe a clear broadening in the $2\theta$ direction even near the epicenter (Fig. 8a), suggesting that the degree of crystal lattice distortion pertaining to the shocked calcite grains was much smaller than in the study of Bell (2016). Shock effects in calcite at peak pressures of 9.0–60.8 GPa were investigated by Bell (2016). The pressure range investigated by Bell (2016) is systematically higher than that of this study (Figure 6a).

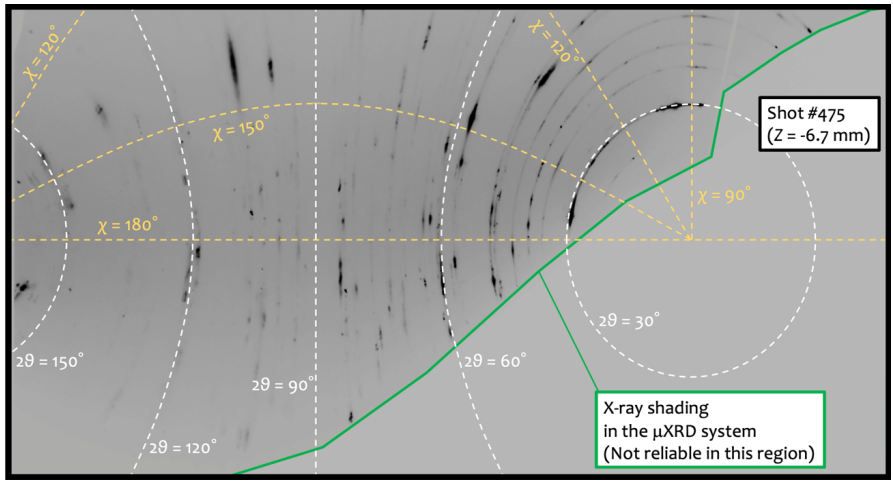

**Figure 7.** Raw two-dimensional XRD data shown as an inverted gray scale image. The data were obtained from the polished thin-section of shot #475. Isolines for the incidence and exit angles ($2\theta$) and another scattering angle ($\chi$) are plotted as dashed lines. XRD data were not obtained in the area surrounded by the green curve because of X-ray shading in the µXRD system.

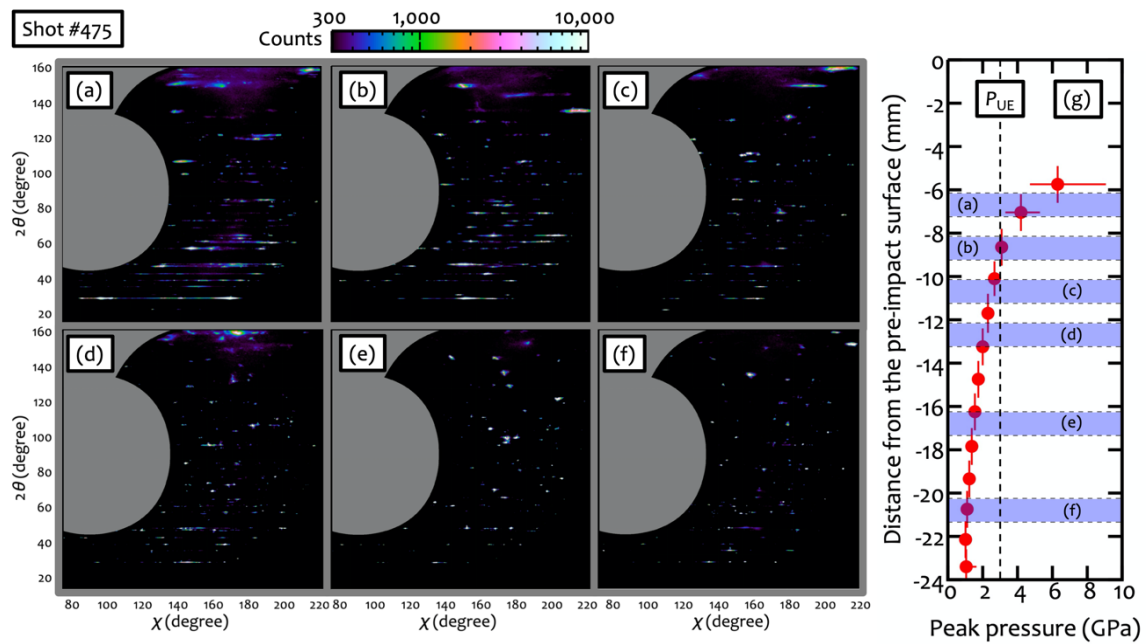

**Figure 8.** (a–f) Re-mapped 2D XRD spectra for shot #475. The horizontal and vertical axes correspond to the angle for another scattering direction ($\chi$) and the sum of the incidence and exit angles of the X-rays ($2\theta$). The observed counts are shown with a color scale in log units. The gray region in each panel is masked because of the X-ray shading as mentioned in Fig. 7. The locations on the polished thin-section of (a–f) are $z = -6.7, -8.7, -10.7, -12.7, -16.8$, and $-20.8$ mm, respectively, where $z$ is the distance from the epicenter in the depth direction. (g) The same as Fig. 4b, except that the corresponding locations for (a–f) are shown as the blue shaded region. The threshold peak pressure for producing undulatory extinction $P_{UE}$ is also shown as the vertical dashed line.

## 4. Discussion

### 4.1. Undulatory extinction as a new shock indicator for calcite grains

Half of the calcite grains in the sub-domains that experienced 2.5–3.0 GPa compression exhibit undulatory extinction. The peak pressure of 2.5–3.0 GPa is slightly higher than the HEL of calcite (1.15–1.42 GPa) [Ahrens and Gregson, 1964], suggesting that the undulatory extinction becomes obvious due to intense plastic deformation of the calcite grains during compressive pulse propagation. It should be mentioned here that the HEL of the macro block of calcite, i.e., Carrara marble, is ~0.1 GPa (See Supplementary Text S4 and Figure S8), which is one order of magnitude lower than the HEL of calcite grain. The observed undulatory extinction indicates that the shocked calcite grains underwent distortion of the crystal lattice by dislocations of a consistent orientation [e.g., Blenkinsop, 2007, pp. 41–43]. Such dislocations in shocked calcite grains have been identified by transmission electron microscopy in previous studies [e.g., Barber and Wenk, 1979; Bell, 2016]. The XRD analyses suggest that a number of dislocations were introduced into the shocked calcite grains because we observed ring patterns in the 2D XRD spectra (Figs. 7 and 8a–b). We propose that the characteristic pressure ($P_{UE}$) for producing undulatory extinction in more than half of the calcite grains is 3 GPa, which is a new shock indicator for calcite at an intermediate peak pressure between the threshold pressures for the production of mechanical twins and incipient devolatilization.

### 4.2. Application to typical catastrophic collisions

If calcite and its carbonate analogues (siderite and dolomite) grains showing undulatory extinction are found in carbonaceous chondrites and return samples from Bennu and Ryugu, this might be used to constrain the origins of the calcite grains, and where water–rock–organic reactions occurred in the parent bodies of carbonaceous chondrites. This is because dynamic compression to 3 GPa can only occur in a region within the distance of a few projectile radii during hypervelocity mutual collisions between asteroids. In this section, we discuss how it is possible to estimate the maximum depth at which calcite grains would have been compressed in their parent bodies based on the presence of undulatory extinction.

We consider a collision that causes a catastrophic disruption of a hypothetical parent body of a group of meteorites. Given that such collisions may be the largest impact events pertaining to the meteorites found on Earth, we may be able to estimate the maximum depth at which the calcite grains were compressed in their parent bodies. We assumed that a projectile with a diameter of 20 km collided with a hypothetical carbonaceous chondrite parent body that had a diameter of 100 km. The impact velocity was assumed to be 5 km s$^{-1}$, which is typical for the main asteroid belt [e.g., Bottke et al., 1994]. This or similar impact conditions, where the size ratio of the impactor to the target is ~0.2 and the impact velocity is 5 km s$^{-1}$, have been widely used to model catastrophic disruptions [e.g., Wakita and Genda, 2019; Michel et al., 2020] because the conditions are statistically plausible. As such, the most likely value of the maximum depth of the shocked calcite grains may be estimated from the undulatory extinction. Figure 9 shows a provenance plot of peak pressure after a typical catastrophic collision in the main asteroid belt. The calculation was conducted in a previous study [Kurosawa et al., 2021b]. The target and impactor materials were modelled with a serpentine EOS with 25% porosity, which is similar to the porosity of CM chondrites [Ostrowski and Bryson, 2019]. Lagrangian tracer particles were used to store the peak pressures depending on their initial location. Further details and the input parameters can be found in the original study [Kurosawa et al., 2021b]. We found that mechanical twins would be produced even at the center of the parent body and that impact devolatilization of calcite occurs only near the impact point. As such, these two shock features are not very useful for reconstructing the depth of the aqueous alteration in the parent bodies of carbonaceous chondrites. In contrast, the isobaric line corresponding to $P_{UE}$ = 3 GPa is located at a depth of three projectile radii from the impact point ($z$ = –30 km). Thus, if calcite-group mineral grains exhibiting undulatory extinction occur in carbonaceous chondrites and/or Ryugu and Bennu return samples, then these calcite grains underwent >3 GPa compression at a depth shallower than 30 km. Furthermore, such calcite grains might have been produced by water–rock–organic reactions in their parent bodies at depths of <30 km, if we assume the impactor had a diameter of <20 km. Although previous and subsequent impact events by a number of smaller impactors might also produce undulatory extinction in calcite grains located near the surface, the location of the isobaric line at 3 GPa at such small impact events must be located nearer the surface.

We assumed that the hypothetical parent body of carbonaceous chondrites is a serpentine body with infinitesimal voids in the shock physics modeling. In contrast, actual carbonaceous chondrites contain a variety of minerals [e.g., Britt et al., 2019] and pores with finite sizes [e.g., Ostrowski and Bryson, 2019]. As such, we need to address the effects of the density contrast between calcite and the surrounding materials [Wittmann et al., 2021] and local pore collapse [Güldemeister et al., 2013] on the peak pressure distributions. Both effects can be neglected in this study because of the following reasons. The minerals surrounding the calcite are likely to be hydrous minerals, because the mass fraction of hydrous minerals in carbonaceous chondrites is several tens of percent [e.g., Britt et al., 2019]. Given that the density of calcite (2.6 Mg m$^{-3}$) [e.g., Pierazzo et al., 1998] is similar to hydrous minerals (~2.5 Mg m$^{-3}$) [e.g., Brookshaw, 1998], the difference in peak pressure between the calcite grains and surrounding materials should be within several percent [Wittmann et al., 2021]. As such, calcite grains are highly suitable shock barometers in carbonaceous asteroids. Güldemeister et al. (2013) demonstrated using shock physics modeling that local pore collapse leads to material convergence into the pores, and a relatively high pressure up to four times greater than the average pressure in the case of a single pore or aligned pores. Güldemeister et al. (2013) also showed that the pressure fluctuations around the average pressure are only ±20% in the case of randomly distributed pores in a shocked media. The random case is more realistic for carbonaceous asteroids. Consequently, the uncertainty in the peak pressure pertaining to the calcite grains in carbonaceous asteroids due to pore collapse would be within a few tens of percent, which does not affect the main conclusions of this study.

It should be also mentioned here that the result with the single impact condition presented in Figure 9 is insufficient to reconstruct the maximum depth of the location of calcite grains. It has been widely known that the peak pressure distributions largely depend on porosity [e.g., Ahrens and O'Keefe, 1972; Wünnemann et al., 2006] and on the impact conditions, including impactor size, impact velocity, and angle [e.g., Pierazzo and Melosh, 2000; Davison et al., 2014]. Further calculations with a range of input parameters are needed. Nevertheless, it would be worthwhile to show the strategy because the shocked calcite grains showing undulatory extinction provide an additional information about the maximum depth of the sample in a multifaceted study that includes the detailed mineralogical, petrological, and chemical analyses of samples, such as meteorites and returned samples from Ryugu and Bennu.

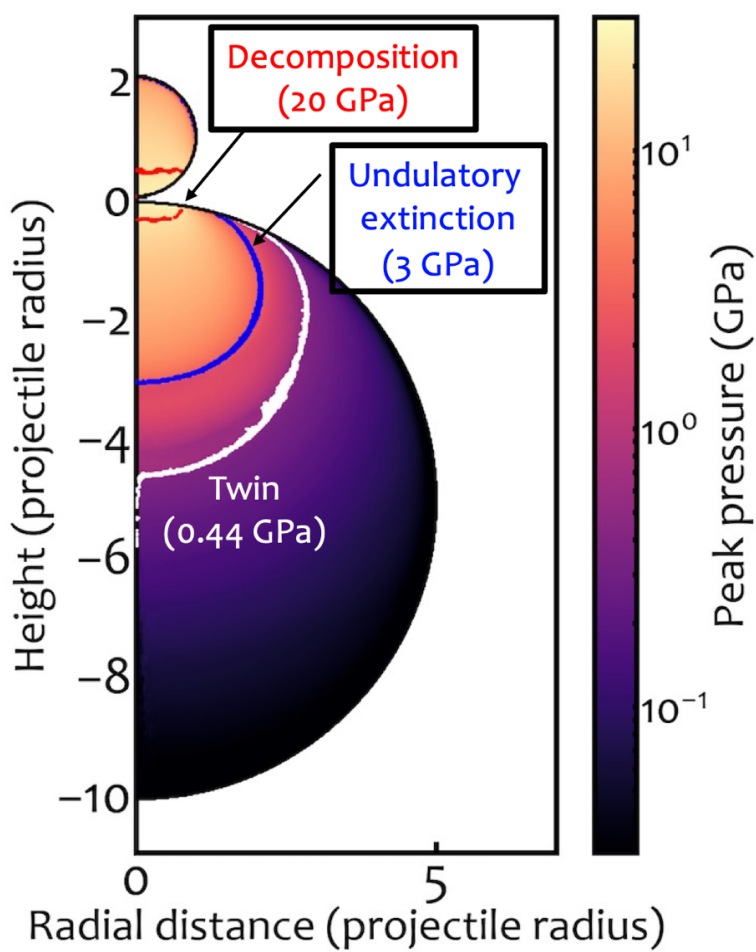

**Figure 9.** Provenance plot of the peak pressure field during a typical catastrophic disruption in the main asteroid belt. Three isobaric lines pertaining to 20 GPa (red), 3 GPa (blue), and 0.44 GPa (white) are shown in the figure. These pressures correspond to the required peak pressures for incipient devolatilization and producing undulatory extinction and mechanical twinning, respectively.

## 5. Conclusions

We investigated shock metamorphism of calcite grains with decaying compressive pulses. This method allowed us to collect shocked areas of marble, which includes calcite grains that experienced a range of peak pressures depending on the distance from the impact epicenter. Shock physics modeling quantified the peak pressure distribution in the shocked samples. We found that the shocked calcite grains that experienced peak pressures of >3 GPa exhibit undulatory extinction under an optical microscope. The threshold pressure required for causing undulatory extinction is slightly higher than the HEL of calcite, indicating that undulatory extinction forms due to intense plastic deformation. We propose that undulatory extinction in calcite and its carbonate analogues is a shock barometer at intermediate shock pressures between those required for producing mechanical twins (~0.5 GPa) and incipient devolatilization (20 GPa). We applied this result to a typical catastrophic collision in the main asteroid belt. If calcite-group mineral grains exhibiting undulatory extinction are found in carbonaceous chondrites and/or return samples from Ryugu and Bennu, these grains are likely to have undergone compression at depths shallower than a few tens of kilometers in their parent body.

**Acknowledgments:** We thank the developers of iSALE, including G. Collins, K. Wünnemann, B. Ivanov, J. Melosh, and D. Elbeshausen. We also thank Tom Davison for the development of pySALEPlot. The authors thank Boris Ivanov for providing the modified parameter set of the ANEOS for calcite. We appreciate two referees for their careful reviews that helped greatly improve the manuscript, and the editor for handling the manuscript. This research was supported by JSPS KAKENHI Grant JP19H00726. K.K. is also supported by JSPS KAKENHI Grants JP18H04464, JP21H01140, and JP21K18660.

**Data Availability Statement**

The iSALE shock physics code is not fully open-source, but is distributed on a case-by-case basis to academic users in the impact community for non-commercial use. A description of the application requirements can be found at the iSALE website (https://isale-code.github.io/terms-of-use.html). The M-ANEOS package is available from Thompson et al. (2019). The list of input parameters for the iSALE computations can be found in the Supplementary Information. The text data supporting the figures in the main text are available in Kurosawa et al. (2021).

527.

**References only in Supplementary Information**

Supporting Information for

**Shock recovery with decaying compressive pulses: Shock effects in calcite (CaCO$_3$) around the Hugoniot elastic limit**

Kosuke Kurosawa[a], Haruka Ono[a], Takafumi Niihara[b,c], Tatsuhiro Sakaiya[d], Tadashi Kondo[d], Naotaka Tomioka[e], Takashi Mikouchi[f], Hidenori Genda[g], Takuya Matsuzaki[h], Masahiro Kayama[i], Mizuho Koike[j], Yuji Sano[h,k], Masafumi Murayama[h], Wataru Satake[l] and Takafumi Matsui[a,l]

[a]Planetary Exploration Research Center, Chiba Institute of Technology, 2-17-1, Tsudanuma, Narashino, Chiba 275-0016, Japan

[b]Department of Systems Innovation, School of Engineering, The University of Tokyo, 7-3-1 Hongo, Bunkyo-ku, Tokyo 113-8656, Japan.

[c]Department of Applied Science, Okayama University of Science, 1-1 Ridaicho, Kita-ku, Okayama, Okayama 113-8656, Japan.

[d]Department of Earth and Space Science, Graduate School of Science, Osaka University, 1-1 Machikaneyama, Toyonaka, Osaka 560-0043, Japan

[e]Kochi Institute for Core Sample Research, Japan Agency for Marine-Earth Science and Technology (JAMSTEC), 200 Monobe Otsu, Nankoku, Kochi 783-8502, Japan.

[f]The University Museum, The University of Tokyo, 7-3-1 Hongo, Bunkyo-ku, Tokyo 113-0033, Japan.

[g]Earth–Life Science Institute, Tokyo Institute of Technology, 2-12-1 Ookayama, Meguro-ku, Tokyo 152-8550, Japan.

[h]Center for Advanced Marine Core Research, Kochi University, 200 Monobe Otsu, Nankoku, Kochi 783-8502, Japan.

[i]Department of General Systems Studies, Graduate School of Arts and Sciences, The University of Tokyo, 3-8-1, Komaba, Meguro-ku, Tokyo 153-8902, Japan

[j]Department of Earth and Planetary Systems Science, Graduate School of Science, Hiroshima University, 1-3-1, Kagamiyama, Higashi–Hiroshima, Hiroshima 739-8526, Japan

[k]Atmosphere and Ocean Research Institute, The University of Tokyo, 5-1-5, Kashiwanoha, Kashiwa, Chiba 277-8564, Japan

[l]Institute of Geo-Cosmology, Chiba Institute of Technology, 2-17-1, Tsudanuma, Narashino, Chiba 275-0016, Japan

*Corresponding author
Kosuke Kurosawa PhD
Planetary Exploration Research Center, Chiba Institute of Technology
E-mail: kosuke.kurosawa@perc.it-chiba.ac.jp

**Contents of this file**

   Text S1 to S6
   Figures S1 to S10
   Tables S1 to S2

**Additional Supporting Information (files uploaded separately)**

   Captions for Movies S1 to S4

**Text S1. Shock physics modeling with the iSALE shock physics code**

We conducted numerical simulations with the iSALE shock physics code [Amsden et al., 1981; Ivanov et al., 1997; Wünnemann et al., 2006; Collins et al., 2016] in order to characterize decaying compressive pulses produced in the impact experiments described in the main text. We have already briefly introduced the numerical calculations in the main text. In this section, we present the parameter set used in the simulations. We used the iSALE "ROCK" model to calculate the elasto-plastic behavior of marble. The ROCK model is a combination of the Lundborg model [Lundborg, 1968] for intact rocks and the Drucker–Prager model [Drucker and Prager, 1952] for damaged rocks. The two models are combined with a damage parameter, which is the ratio of the total plastic strain to the threshold volumetric strain [Ivanov et al., 1997]. We also used a thermal softening model [Ohnaka, 1995]. The input parameters for the material models, including the Equations of State (EOS) and strength models, are summarized in Table S1. The calculation settings are described in Table S2. We used pySALEPlot software to analyze and visualize the calculated results.

We used the von Mises strength model pertaining to titanium. The limiting strength $Y_{\text{lim}}$ is the required parameter for the model. We determined $Y_{\text{lim}}$ to be 0.5 GPa with a trial-and-error method by comparison with the resultant impact crater on a detachable front plate. Figure S1 shows the numerical results with three different $Y_{\text{lim}}$ values of 0.05, 0.5, and 5 GPa. The crater size on the detachable front plate at $Y_{\text{lim}} = 0.5$ GPa is consistent with the experiment. It was not important whether the front plate was eventually penetrated, as the compression wave propagated into the target much earlier than penetration of the plate. The materials of the front plate that were initially located beneath the projectile footprint became deformed and thinner than the initial thickness of the front plate (Figure S1b). A higher spatial resolution would be required to investigate whether the front plate would eventually tear off, but such calculations were beyond the scope of this study.

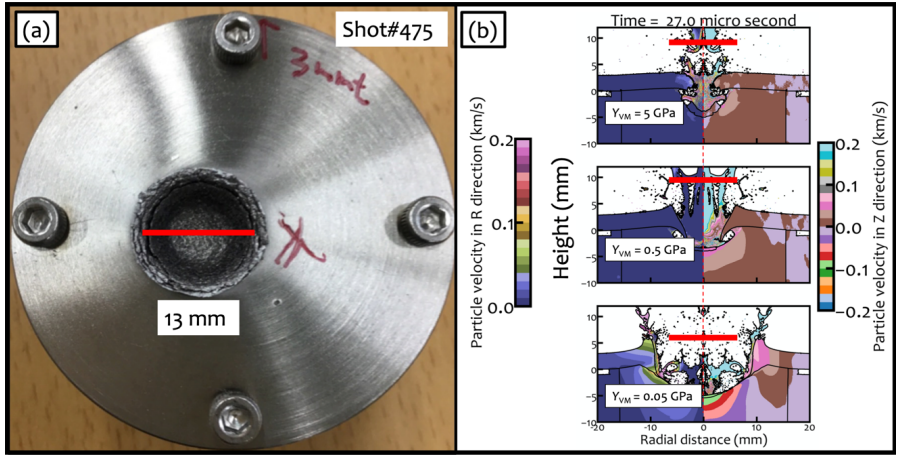

**Figure S1.** (a) An impact crater on a detachable front plate. The crater diameter was measured to be 13 mm as shown as the red horizontal bar. (b) Effects of the von Mises strength $Y_{VM}$ of titanium on the crater size on the front plate. The color scales correspond to the particle velocities in the radial (left) and depth (right) directions. The red horizontal bar is the crater diameter determined in the panel (a). At this time, the crater growth in the case with $Y_{VM}$ = 0.5 GPa mostly ceased, and the resultant crater size is consistent with the crater diameter in shot #475.

**Table S1.** Input parameters used in the iSALE simulations. The details of each parameter can be found in the iSALE manual [Collins et al., 2016].

|  | Polycarbonate projectile | Al plate | Ti plate & container | Marble |
|---|---|---|---|---|
| EOS model | Tillotson EOS[a,b] | Tillotson EOS[a] | Tillotson EOS[a] | ANEOS calcite[c,d] |
| Strength model | HYDRO | JNCK[e] | VNMS | ROCK IVANOV[f] |
| Poisson ratio | 0.5 | 0.33[g] | 0.34[h] | 0.3[i] |
| Minimum pressure (GPa) | Not used | -2.5 | -0.5 | Not used |
| Melting temperature (K) | Not used | 978[j] | 1,941[k] | 1,690[l] |
| Specific heat (J K$^{-1}$ kg$^{-1}$) | Not used | 896[g] | 521[m] | Calculated in ANEOS package |
| Thermal softening coefficient | Not used | Not used | Not used | 1.2[i] |
| Simon parameter $a$ (GPa) | Not used | 7.98[j] | 7.0[k] | 25.3[n] |
| Simon parameter $c$ | Not used | 0.57[j] | 7.19[k] | 2.26[n] |
| Cohesion (intact) (MPa) | Not used | Not used | Not used | 50[i] |
| Cohesion (Damaged) (MPa) | Not used | Not used | Not used | 0.1[i] |
| Internal friction (intact) | Not used | Not used | Not used | 2[i] |
| Internal friction (Damaged) | Not used | Not used | Not used | 0.4[i] |
| Limiting strength (GPa) | Not used | Not used | 0.5[o] | 2.6[n] |

| Minimum failure strain | Not used | Not used | Not used | $10^{-4}$ [p] |
|---|---|---|---|---|
| Constant for the damage model | Not used | Not used | Not used | $10^{-11}$ [p] |
| Threshold pressure for the damage model (MPa) | Not used | Not used | Not used | 300[p] |
| Johnson-Cook parameter $A$ (MPa) | Not used | 324[q] | Not used | Not used |
| Johnson-Cook parameter $B$ (MPa) | Not used | 114[q] | Not used | Not used |
| Johnson-Cook parameter $N$ | Not used | 0.42[q] | Not used | Not used |
| Johnson-Cook parameter $C$ | Not used | $2 \times 10^{-3}$ [q] | Not used | Not used |
| Johnson-Cook parameter $M$ | Not used | 1.34[q] | Not used | Not used |
| Johnson-Cook parameter $T_{\mathrm{ref}}$ (K) | Not used | 298 | Not used | Not used |

a. Tillotson [1962].

b. The Tillotson parameters for polycarbonate were obtained by Sugita and Schultz [2003], based on the shock Hugoniot data compiled by Marsh [1980].

c. Thompson and Lauson [1972].

d. The ANEOS parameters for calcite were determined by Pierazzo et al. [1998]. Subsequently, the parameter set was slightly modified by the impact community. The modified values were listed in Kurosawa et al. [2021].

e. Johnson and Cook [1983].

f. Ivanov et al. [1997].

g. We used an aluminum alloy (Al-6061). The Poisson ratio was taken from the data compiled on the web page (http://asm.matweb.com/search/GetReference.asp?bassnum=ma6061t6) provided by ASM Aerospace Speciation Metals Inc. The database was constructed based on the information provided by the Aluminum Association Inc.

h. Pure titanium was used for the front plates and metal containers. The Poisson's ratio was taken from the data compiled by Köster and Franz [1961].

i. Goldin et al. [2006].

j. Hänström and Lazor [2000].

k. Kulyamina et al. [2018].

l. The lowest melting temperature on the pressure–temperature plane (3 MPa) from Kerley [1989] was used for the melting temperature under the reference state.

m. The Dulong–Petit limit was used.

n. The melting curve used in the simulation was approximated with the Simon equations [e.g., Wünnemann et al., 2008]. The fitting was conducted by Kurosawa et al. (2021) with the melting curve of Kerley [1989].

o. See the Supplementary Text S1.

p. Typical values for rocks were used.

q. Rule [1997].

**Table S2.** Calculation settings used in the iSALE simulations. A detailed description of each parameter can be found in the iSALE manual [Collins et al., 2016].

| | |
|---|---|
| Computational geometry | Cylindrical coordinates |
| Number of computational cells in the $R$ direction | 700 |
| Number of computational cells in the $Z$ direction | 1,200 |
| Grid spacing (mm/cell) | 0.05 |
| Cells per projectile radius (CPPR) | 46 |
| Artificial viscosity $a_1$ | 0.24 |
| Artificial viscosity $a_2$ | 1.2 |
| Impact velocity (km s$^{-1}$) | Same as for the impact experiments |
| Initial temperature (K) | 300 |
| High-speed cutoff | Double the impact velocity |
| Low-density cutoff (kg m$^{-3}$) | 1 |

**Text S2. Resolution effect**

It has been widely known that the spatial resolution largely affects the outcomes of shock physics modeling [e.g., Pierazzo et al., 2008]. The number of cells per projectile radius ($n_{CPPR}$) is one parameter that determines the spatial resolution in numerical simulations. According to Pierazzo et al. (2008), $n_{CPPR}$ should be >20 to accurately estimate the peak pressure distribution. Although we chose $n_{CPPR}$ = 46, as mentioned in the main text and listed in Table S2, we also investigated the effects of resolution on the impact modeling. We conducted two additional runs pertaining to shot #475 with $n_{CPPR}$ = 23 and 92, which correspond to half and double the values of $n_{CPPR}$ used in this study, respectively. The other impact conditions were the same as used for Figures 2, 3, and 4b. Figure S2 is the same as Figure 2 in the main text, except that the effects of $n_{CPPR}$ on the propagation behavior of a compressive pulse are shown. We confirmed that the results are mostly the same irrespective of $n_{CPPR}$, except for the strain rate. The strain rate became higher with increasing $n_{CPPR}$. The strain rate at the wave front in the case for $n_{CPPR}$ = 92 is about two times that obtained with the fiducial value ($n_{CPPR}$ = 46). Given that the strain rate is one of the basic physical quantities that characterizes shock recovery experiments [e.g., Kimberley et al., 2013; Rae et al., 2022], we presented an order of magnitude estimate of strain rate under our experimental conditions. Further investigations of the effects of $n_{CPPR}$ on the strain rate will be needed to use the strain rate for more advanced analyses [e.g., Wiggins et al., 2019; Rae et al., 2022].

The most important variable obtained by the shock physics modeling in this study is the mass-weighted average values of the peak pressures in the sub-domains, because these values are used to obtain the characteristic pressure ($P_{UE}$) for producing undulatory extinction in more than half the calcite grains. Figure

S3 shows the peak pressures obtained from the fiducial results with $n_{CPPR} = 46$ and with $n_{CPPR} = 23$ and 92. We confirmed that the peak pressures are similar irrespective of the $n_{CPPR}$ value. As such, our choice of $n_{CPPR} = 46$ was sufficiently high to accurately estimate the peak pressure distributions in the shocked targets.

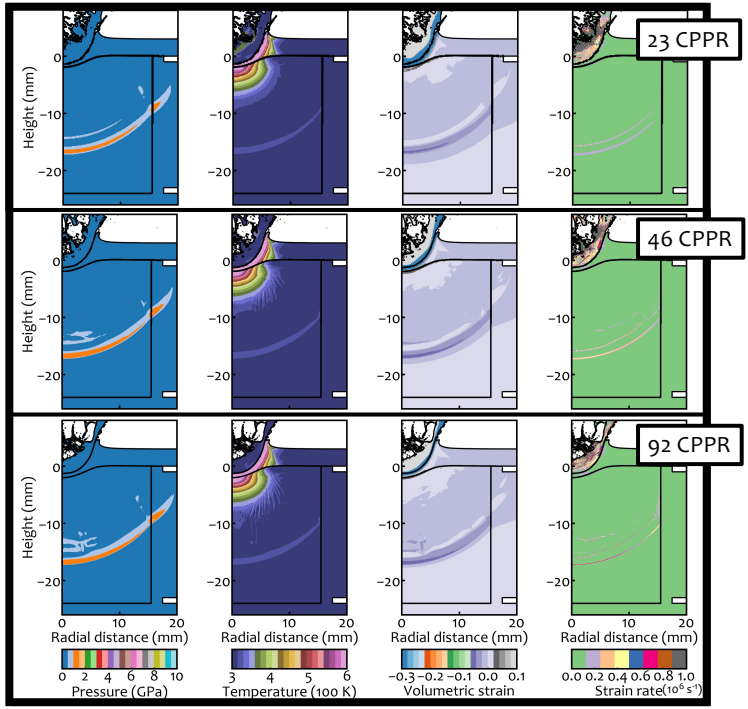

**Figure S2.** Same as Figure 2 in the main text, but showing the effects of the number of cells per projectile radius on the pressure, temperature, volumetric strain, and strain rate at 4 μs after the collision.

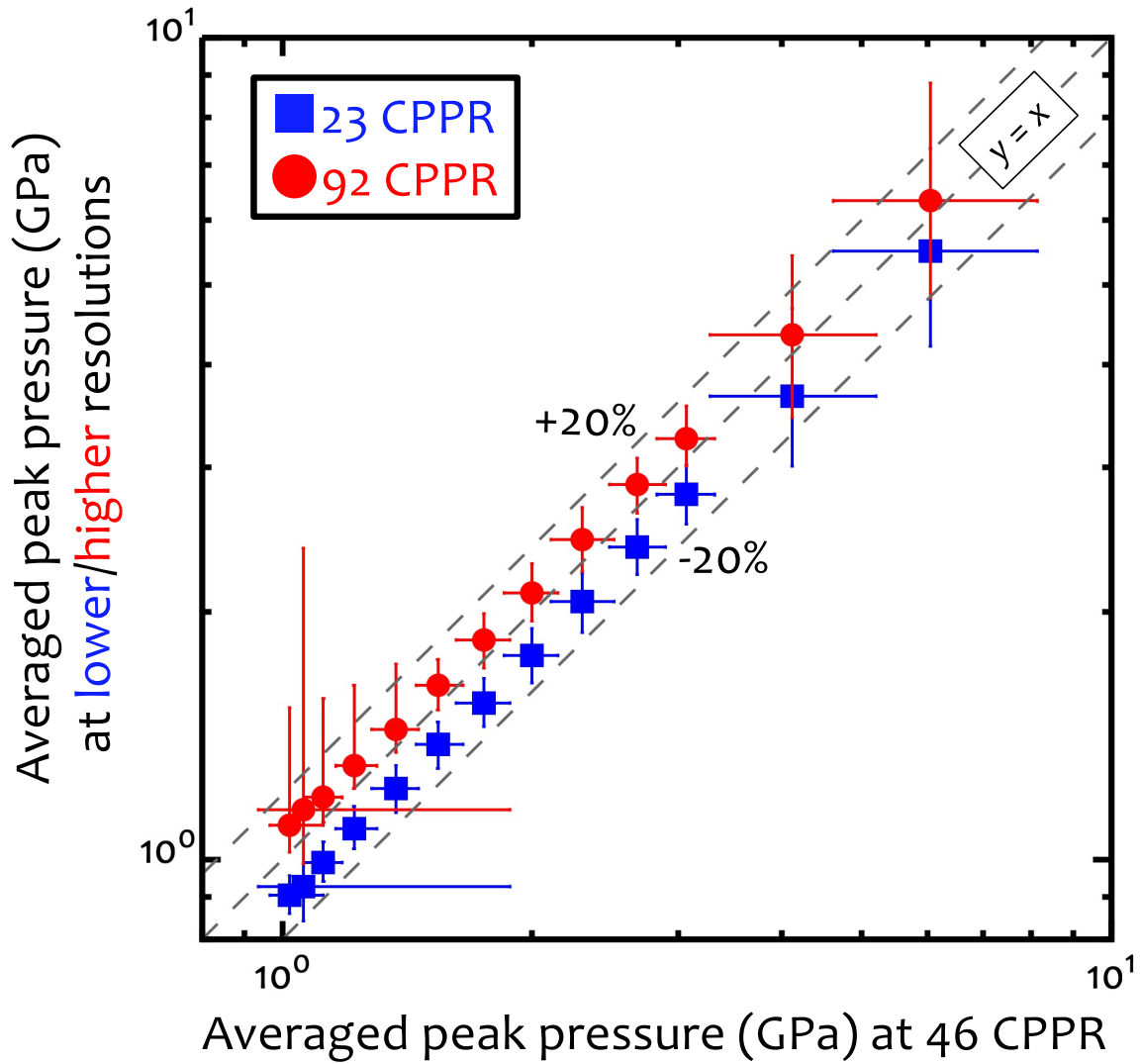

**Figure S3.** Effect of the number of cells of the projectile radius on the mass-weighted average values of the peak pressures in the sub-domains. The gray dashed lines indicate the 1:1 correlation and uncertainties of ±20%.

**Text S3. Effects of the strength parameters of marble**

In this section, we discuss the uncertainties on the input parameters of the ROCK model pertaining to marble. Although we used the parameter set of Kurosawa et al. (2021a), which is basically the same as in Goldin et al. (2006), except for the value of the limiting strength, Winkler et al. (2018) presented a different set of parameters for marble. As such, we investigated the effects of the parameter sets on the impact modeling by conducting an additional run using the parameter set of Winkler et al. (2018), as described in Text S2. Figures S4 and S5 are the same as Figures S2 and S3, except that the results obtained with the parameters listed in Table S1 and those of Winkler et al. (2018) are shown, respectively. The iSALE calculations with two different sets of parameters yielded a similar propagation behavior for the compressive pulse (Figure S4). We confirmed that the peak pressures are largely similar, except for the pressure range from 2 to 3 GPa (Figure S5). The parameter set of Winkler et al. (2018) tends to yield somewhat lower peak pressures in this pressure range, although the difference is <20%. This small difference does not change the values of $P_{UE}$.

Justo and Castro (2021) reported that the Poisson's ratio of Carrara marble is 0.316–0.386. In contrast, we selected a Poisson's ratio of 0.300, which is close to the lowest value. As such, we also conducted an additional run with the highest value (0.386) of Poisson's ratio. Figures S6 and S7 are the same as Figures S4 and S5, except that the results for Poisson's ratios of 0.300 and 0.386 are compared. Figure S6 shows that the location of the wave front and strength of the compressive pulse, in the case with a Poisson's ratio of 0.386, are similar to the case with a Poisson's ratio of 0.300. A Poisson's ratio of 0.386 tends to yield higher peak pressures over the entire range of investigated pressures, but the difference is within 40% (Figure S7). This small change does not significantly change the values of $P_{UE}$.

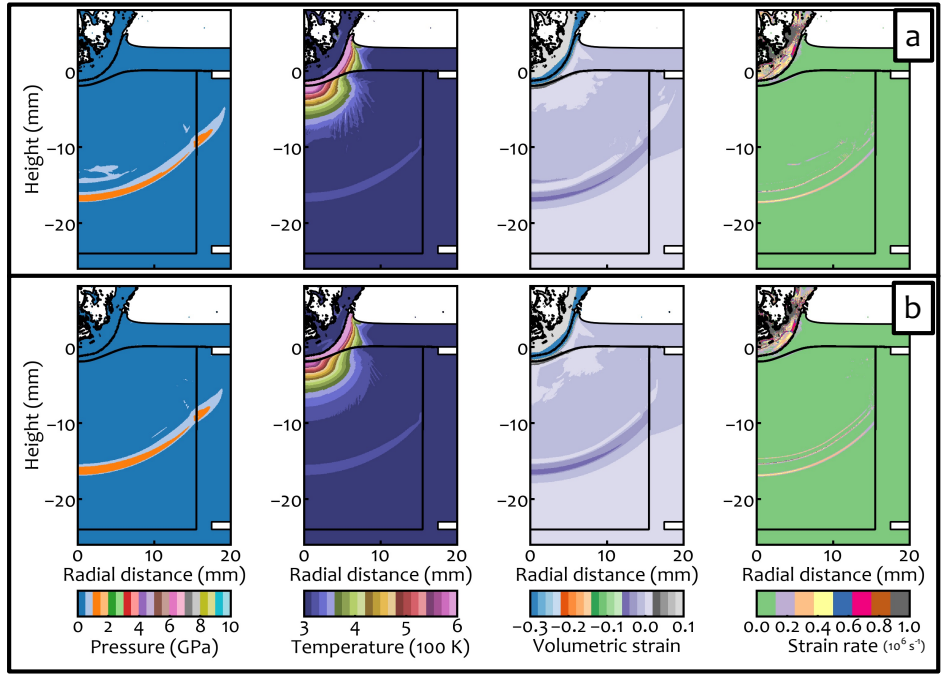

**Figure S4.** Same as Figure S2, but showing the effects of the selection of the strength parameters pertaining to marble. The results were obtained with the parameters (a) listed in Table S1 and (b) of Winkler et al. (2018).

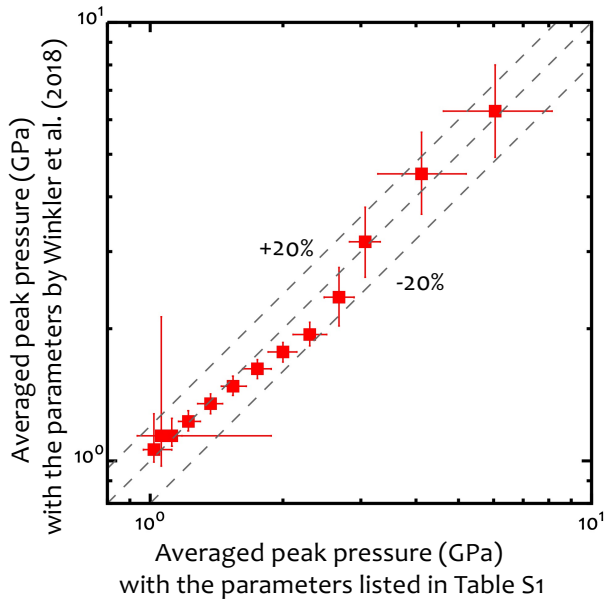

**Figure S4.** The same as Figure S3, except that the correlation between the fiducial results and the results with the different set of the strength parameters for marble is shown.

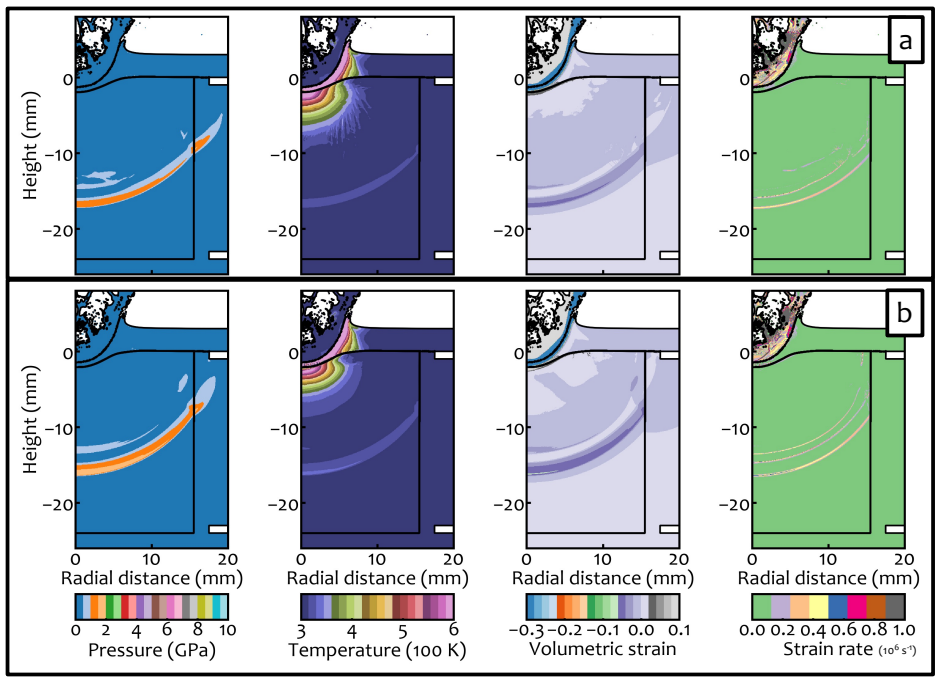

**Figure S6.** Same as Figures S2 and S4, but showing the effects of the Poisson's ratio. The results are shown for Poisson's ratios of (a) 0.300 and (b) 0.386.

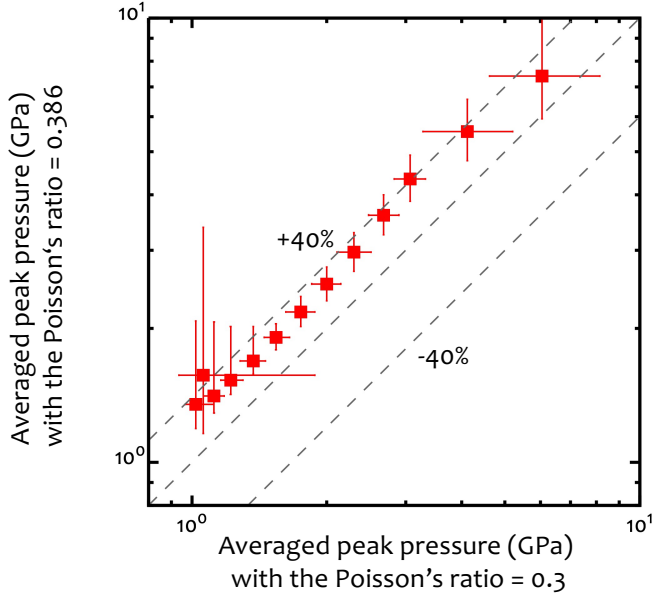

**Figure S7.** Same as Figures S3 and S5, but showing the correlation between the fiducial results and the results with different values of Poisson's ratio. Uncertainties of ±40% are shown as gray dashed lines.

**Text S4. Stress measurements and comparison with the iSALE results**

We conducted direct stress measurements for shots #486 and #487 as described in the main text to verify the validity of the iSALE results. We used a PolyVinylidene DiFluoride (PVDF) film gauge (PVF2 11-.125-EK Uni, Dynasen Inc.) and a charge integrator (CI-50-0.1, Dynasen Inc.) to measure the normal stress during impacts. The dimension of the sensor is $3.2 \times 3.2$ mm. The thickness of the gauge including the film is 50 μm. The two film gauges were inserted along the boundaries between the front/rear detachable plates and marble specimen. During these measurements, we used aluminum instead of titanium, because the shock impedance of aluminum is similar to that of marble. Given that a number of wave reflections through the PVDF gauge are expected to occur within a very short time scale, which is much shorter than the characteristic time for projectile penetration, we can measure the normal stress parallel to the projectile trajectory in the marble at the epicenter and on its rear surface.

We only analyzed the data obtained for shot #487 because the projectile collided on the aluminum front plate at the specific location where it is directly above the sensor surface. Thus, we were able to measure the maximum principal stress at the epicenter and rear surface, which can be compared with the iSALE simulations. The maximum principal stress $\sigma_L$ is given by Poisson's ratio $\nu$ and pressure in the iSALE $P_{iSALE}$ as [e.g., Melosh, 1989, pp. 34]:

$$\sigma_L = \frac{3(1-\nu)}{1+\nu} P_{iSALE}. \text{ (Eq. S1)}$$

The maximum principal stress $\sigma_L$ is 1.62 times $P_{iSALE}$. We extracted the temporal variation of $P_{iSALE}$ using the tracer particles located initially at $(r, z)$ = (1.0 mm, 0.0 mm) and (1.0 mm, –24.0 mm) in the simulation, where $r$ and $z$ are the radial distance and height from the epicenter, respectively. We chose $r$ = 1.0 mm, which is 20 cells away from the symmetry axis, because $r$ = 0.0 mm is too close to the symmetry axis in the 2D simulation. Figure S8 shows the changes in $\sigma_L$ at the epicenter and rear surface of the marble as a function of time. The iSALE results corresponding to the front and rear gauges are also shown in the figure. The upper and lower lines for the iSALE results are $\sigma_L$ and $P_{iSALE}$, respectively. Note that the front gauge was broken after the wave arrival. Thus, the time profile from the front gauge is unreliable. Nevertheless, the peak value of $\sigma_L$ at the epicenter can be used to verify the iSALE result. We found that the peak value of $\sigma_L$ calculated by iSALE pertaining to the front gauge was entirely consistent with the direct stress measurement. In contrast, the data for the rear surface are quite different. Although wave splitting was detected by the rear gauge, our iSALE model did not reproduce such behavior. This is possibly because the treatment of the relationship between the longitudinal stress and wave propagation speed in the current version of the ANEOS calcite is too simplified. Figure S9 shows the wave propagation speed as a function of the particle velocity for marble [Pierazzo et al., 1998, and references therein] along with the prediction by the ANEOS calcite. In general, the longitudinal sound speed $C_L$ is faster than the bulk sound speed $C_B$. The value of $C_L$ for the Carrara marble is reported to be 6.2 km s$^{-1}$ [e.g., Piementa et al., 2018]. In contrast, the ANEOS calcite yields $C_B$ of 3.35 km s$^{-1}$ (the intercept of the $y$-axis in Figure S9) and reproduces the shock Hugoniot data. The elastic waves, which cause the longitudinal stress to be lower than the Hugoniot elastic limit, propagate in media at $C_L$, although the propagation speeds of the plastic and shock waves follow the prediction of the ANOES calcite. Figure S8 also shows that the Hugoniot elastic limit of the Carrara marble used in this study is 0.09 GPa, as is evident from an abrupt increase in the stress value at 4.4 μs recorded by the rear gauge. Given that there is a speed difference between $C_L$ and $C_B$ of up to 2.85 km s$^{-1}$ in the case of Carrara marble, an elastic wave precedes the plastic wave front. Such wave splitting is known to be an elastic precursor. Wave splitting was not reproduced in the iSALE simulation with the current version of the ANEOS calcite, resulting in momentum transport with a longer dwell time, such that the peak $\sigma_L$ value from the iSALE model is an overestimation (Figure S8). The time difference between the arrivals of the elastic and plastic compressive pulses becomes larger with increasing distance from the epicenter due to the speed difference between the elastic and plastic waves. Therefore, the degree of overestimation of $\sigma_L$ is also likely to increase with distance. However, the difference in $\sigma_L$ in the target between the experiment and simulation is within a factor of two, even at the rear surface of the target, where the difference is expected to be the largest. This difference does not significantly affect the $P_{UE}$ measurements. Consequently, the iSALE model used in this study provides peak pressures and stress distributions that are of sufficient accuracy.

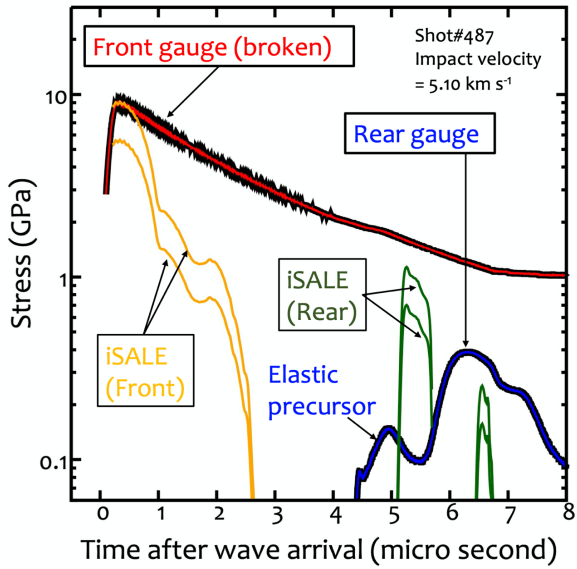

**Figure S8.** Time variations of the maximum principal stresses at the epicenter and the rear surface of the target (red and blue lines). The time origin was adjusted to the arrival time of the compressive pulse to the front gauge. The iSALE results for the epicenter (front, orange) and the rear surface (rear, green) are also shown. The Hugoniot elastic limit of the Carrara marble can be measured to be 0.09 GPa with the data by the rear gauge as shown as the horizontal black arrow.

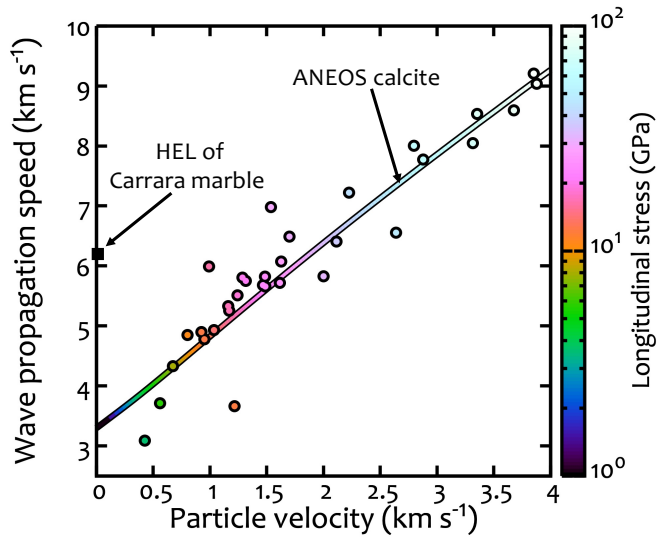

**Figure S9.** Wave propagation speed in marble as a function of particle velocity. The prediction of the ANEOS calcite is shown as the colored line, and filled circles represent experimental data compiled by Pierazzo et al. (1998). The colors correspond to the longitudinal stress. The Hugoniot elastic limit (HEL) of marble is shown as the black square.

**Text S5. Three-dimensional X-ray computational tomography of a shocked sample**

We observed the three-dimensional damage structure of the recovered sample from shot #475 with a micro-focus X-ray computational tomography (XCT) device (Zeiss Xradia 410 Versa) at Kochi University, Japan, prior to making the polished thin-section as mentioned in the main text. The system has an X-ray tube with a tungsten target. Given that we scanned the entire recovered sample with a diameter of 30 mm, the spatial resolution was limited to ~33 μm. We used collecting optics with a 0.4× objective. The X-ray tube voltage and power were set to 140 kV and 8.0 W, respectively. The exposure time and number of projections were set to 2 s and 1,601, respectively. The rotation angle was 360°. Figure S10 shows examples of 3D XCT images. In this figure, we present typical cross-sections across the center of the recovered sample. We observed a number of cracks with widths of >100 μm, and we confirmed that there is no clear dependence of the azimuth angle on the damage structures. Although we only observed a single specific cross-section of the shocked samples by polarizing microscopy and X-ray diffraction analysis, as described in the main text, it does not change the main conclusion of this study. Here we demonstrated that we can observe the 3D damage structure of a shocked sample that has experienced a decaying compressive pulse. Better spatial resolution could be obtained if the recovered sample was a cylindrical shape and had a smaller diameter. Such advanced observations will be conducted in a future study.

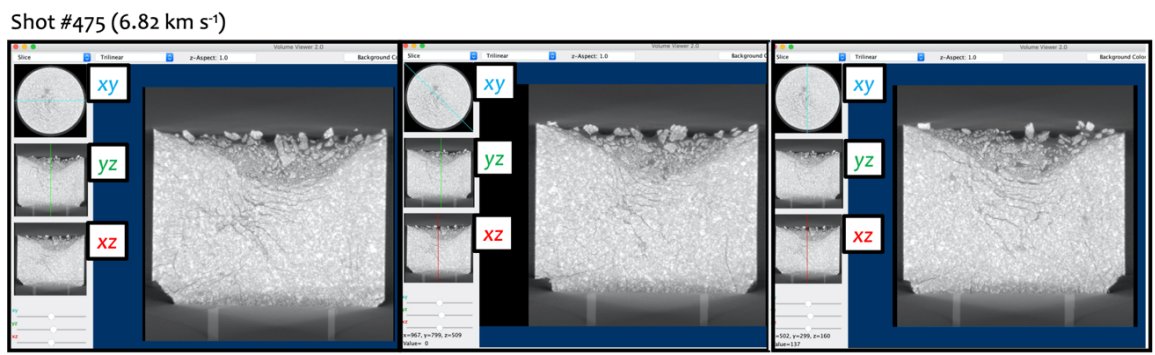

**Figure S10.** Three-dimensional X-ray computational tomography (XCT) images for shot #475. Three typical cross-sections across the center of the shocked marble cylinder are shown. The cross-section planes are shown as cyan, green, and red lines on the three small images.

**Text S6. Microscopic observations of polished thin-sections**

We present several examples of the observations with a polarizing microscope as Supplementary Movies. We observed the changes in the extinction pattern of calcite grains as the table on the polarizing microscope was rotated. Although we only show movies taken of the polished thin-section from shot #487, the extinction behavior is similar for the other shots.

**Captions for Supplementary Movies**

**Movie S1.** Transmitted light image in cross-polarized light taken at $z = -1.7$ mm, where $z$ is the distance from the epicenter in the depth direction. The corresponding peak pressure estimated by the iSALE $P_{peak}$ value is 5.5 GPa.

**Movie S2.** The same as Movie S1, except that $z = -5.8$ mm and $P_{peak} = 2.5$ GPa.

**Movie S3.** The same as Movie S1, except that $z = -15$ mm and $P_{peak} = 1.2$ GPa.

**Movie S4.** The same as Movie S1, except that $z = -20$ mm and $P_{peak} = 0.83$ GPa.